\documentclass[useAMS,usenatbib]{mnras}
 \usepackage{graphicx,caption}
 \usepackage{graphics}
 \usepackage{multicol}
 \usepackage{subfig}
 \usepackage{rotating}
 \usepackage{etex}
 \usepackage{setspace}
 \usepackage{float}
\usepackage[american]{babel}
\usepackage[T1]{fontenc}
\usepackage{ae,aecompl}
\usepackage{hyperref}
\usepackage{tabularx}
\usepackage{pdflscape}
\usepackage{natbib}
\usepackage{longtable}

\title[Spectral study of MAXI J1957+032]{Unveiling the nature of compact object in the LMXB MAXI~J1957+032 using \emph{Swift}-\textsc{XRT} }
\author[A. Beri et al.]
  {
  Aru~Beri$^{1,2}$ \thanks{a.beri@soton.ac.uk},
  D.~Altamirano$^2$, 
   R.~Wijnands$^3$, 
   N.~Degenaar$^3$, 
   A.S.~Parikh$^3$, 
   K. Yamaoka$^4$ \\
 $^1$, DST-INSPIRE Faculty, Indian Institute of Science Education and Research (IISER) Mohali, 
Punjab 140306, India  \\ 
 $^2$, School of Physics and Astronomy, University of Southampton,~Southampton,~Hampshire SO17~1BJ,~UK \\
 $^3$, Anton Pannekoek Institute for Astronomy, University of Amsterdam, Postbus 94249, 1090 GE Amsterdam, The Netherlands \\
 $^4$, Nagoya University, Furo, Chikusa, Nagoya 464-8601, Japan \\
 }

\begin{document}
\pagerange{\pageref{firstpage}--\pageref{lastpage}} 
\maketitle
\label{firstpage}
\begin{abstract}

  MAXI~J1957+032 is a transient Low-mass X-ray binary (LMXB) that
  underwent four short outbursts in 1.5 years since its discovery in
  2015. The nature of the compact object in MAXI~J1957+032 is not clear,
  but it was proposed to be a neutron star based on the short-duration
  of its outbursts.
  Here, we report the results obtained after performing spectral
  analysis using data obtained with the X-ray telescope aboard the Neil Gehrels \emph{Swift} satellite.~When
  describing the spectrum with an \textit{absorbed power law}, we
  found that the spectra softens~(the power-law index increases from $\sim$ 1.8 to 2.5) as the luminosity decreases. 
  Near the end of its outbursts the observed value of power-law index~($\Gamma$) is
  $\sim$~2.5.
   To identify the nature of the compact object in this system,
  we used $\Gamma$ as a tracer of the spectral evolution with
  luminosity. 
 We found that for the distance of 4~kpc, 
  our results suggest that the source harbours a neutron star.

\end{abstract}

\section{Introduction}
Low mass X-ray binaries~(LMXBs) contain a black hole~(BH) or neutron star~(NS)
that accretes matter from a companion star which typically has a mass
lower than that of the accretor.  Most low mass X-ray binaries are
transients which means they undergo outbursts sporadically while
spending most of their time in a quiescent state with X-ray luminosity
below $\sim10^{34}$ $\mathrm{ergs~s^{-1}}$.  During an outburst, the X-ray luminosity
can increase up to a few times {\bf $10^{36}-10^{39}$} $\mathrm{ergs~s^{-1}}$.
In the last 15 years, it has been found that there are LMXBs which show
sub-luminous accretion outbursts, i.e., having peak outburst luminosities within a
range of $10^{34}-10^{36}$ $\mathrm{ergs~s^{-1}}$.~This class of LMXBs is known as very faint
X-ray transients~\citep[VFXTs; see, e.g., ][]{Wijnands06}.

One of the challenging aspects in the study of LMXB -- and in
particular of these VFXTs -- is to understand
the nature of a compact object.
There are only a few observational methods to unambiguously constrain
it. If the source shows coherent
pulsations or thermonuclear X-ray bursts, then it is a NS. If the mass
function (typically from optical/infrared spectroscopy) is measured,
then one can generally state if the binary system hosts a NS or a 
BH.~In cases where we lack these measurements, the possible nature of 
its compact object is inferred from the comparison of its spectral and timing signatures
with those observed
from systems for which we know the nature of their compact objects \citep[see e.g.][]{Porquet05}.
Studies carried out during the quiescent state can also be used to infer the nature of the accretor 
\citep[see e.g.][]{Gelino06}.

The faint luminosities of VFXTs make them difficult 
to find with all-sky monitors whose sensitivity is typically not
sufficient to detect those systems beyond the peak of their outburst.
When followed-up and monitored with more sensitive instruments, especially 
with \emph{Swift}, the VFXTs often appear to show brief (few days to few weeks) outbursts,
hence allowing only for a limited time opportunity to study these systems
\citep[e.g.][]{Armas11,Armas14,Degenaar15}.
In addition, many are found in the Galactic plane and have 
thus large distances and typically have large column densities
toward them. This does not allow for detailed studies of their 
optical or near-infrared properties and hence not much is 
known about their companion stars either. \\

MAXI~J1957+032~(also sometimes referred to as IGR~J19566+0326) is an X-ray binary which was first observed
in outburst on May 11, 2015 with the Gas Slit Camera~(\textsc{GSC})
aboard \emph{MAXI} \citep{Negoro15}. It was later also detected using
\emph{Integral} in 20-60 keV band \citep{Cherepashchuk15}.
\emph{Chandra} also observed the source giving the best known position to date \citep{Chakrabarty16}.
Since its discovery, MAXI~J1957+032 exhibited four outbursts; each outburst decayed quite rapidly \citep[within a few
  days, see eg., ][]{Mata17}.
%
These authors showed that the optical spectrum of MAXI~J1957+032 during its outburst is
consistent with other LMXB transients.~The
same authors also proposed that the source is a NS~system
based on the resemblance of the system with the 
accreting millisecond X-ray pulsar~(AMXP) in NGC~6440~X--2.
Both systems have a exhibited a period during which they displayed
frequent outbursts which only lasted several days.~For a reasonable range~(2--8~kpc) of distances, MAXI~J1957+032 classifies as a VFXT. \\

So far, no BH system has been identified that also {\bf exhibits} such short, frequent outbursts.
However, it is unclear whether short outbursts are characteristic only for neutron star
X-ray binaries or that they can also indeed be observed in BH systems. 
\citet{Knevitt14} argued that for BHs with short orbital periods ($P_{orb}{<} 4h$),
the peak outburst luminosity drops close to the
threshold for radiatively inefficient accretion, and BH LMXBs
can have lower outbursts luminosities and shorter outburst durations
compared to NS systems.~However, there exist systems
like Swift~J1357.2-0933 \citep[see e.g.,][]{MataS15, Armas13} and Swift~J1753.5-0127 \citep[see e.g.,][]{Ramadevi07, Zurita08} 
that are believed
to have short orbital periods, but they do not exhibit short outbursts.
Moreover, MAXI~J1659--152 is a short orbital period BH but it exhibits
bright outbursts \citep[][]{Kuulkers13}. \\
 
\citet{Wijnands15} searched the literature for reports on the spectral properties of NS and BH LMXBs
studied using an absorbed \textit{power-law} model.
They compared the spectra of NS and BH transients 
when they have accretion luminosities between $10^{34}-10^{36}$~$\rm{ergs~s^{-1}}$.
The authors found that NSs are significantly softer than BHs below an
X-ray luminosity~(0.5-10~keV) of $10^{35}$ $\mathrm{ergs~s^{-1}}$ \citep[Figure~1
  of][]{Wijnands15}.~Thus, the spectral shape between 0.5 and 10~keV
at these low luminosities can be a useful tool to distinguish LMXBs
hosting NSs from those harbouring BHs.
In this paper we present the evolution of all the outbursts observed
with MAXI and Swift, and we used the results obtained by \citet{Wijnands15} to try
to get more insights into the nature of MAXI~J1957+032.

 \section{Observations}

The Neil Gehrels \emph{Swift} observatory, launched in November 2004 \citep{Gehrels04}, has 
three instruments on board:~a) the Burst Alert Telescope
(\textsc{BAT}) which operates in the energy range of 15-150 keV
\citep{Barthelmy05}; b) the X-ray Telescope (\textsc{XRT}), operating in
the range of 0.2-10~keV \citep{Burrows07}; c) the Ultraviolet and Optical
Telescope (\textsc{UVOT}) which covers UV and optical bands
\citep[170--600~nm;][]{Roming04}.~In our paper, we have used the
observations of MAXI~J1957+032 performed with \emph{Swift}-\textsc{XRT}
during the 2015 and 2016 outbursts of the source.~The 
details of observations used are given in Table~\ref{Swift}.
Due to the rapid decay of the source during the outbursts, 
only during 8 observations enough photons were detected to allow for meaningful
spectral analysis to be carried out. \\

\begin{table*}
\caption{Log of observations made with \emph{Swift}-\textsc{XRT} in response to the outbursts of MAXI~J1957+032.
All errors and upper limits are at 3-$\sigma$ confidence level.}
\label{table2}
\begin{tabular}{ c c c c c c c c c}
\hline
\hline
Obs-ID & Time~(MJD) & Mode  & Exp-time~(ksec) & 0.3-10~keV count rate~($\rm{count~s^{-1}}$)   \\
\hline
\hline
33770001 & 57156.03  & PC    & 3.0 & $0.55\pm0.02$   \\
33770002 & 57157.67  & PC    & 2.0 & $0.013\pm0.003$ \\
33770005 & 57162.06  & PC    & 2.8 & $0.006\pm0.001$  \\ 
33770006 & 57164.76  & PC    & 0.9 &$0.008\pm0.004$   \\
33770007 & 57165.78  & PC    & 2.2 & $0.006\pm0.002$  \\
33770009 & 57304.68  & PC    & 1.0 & $0.65\pm0.04$ \\
33770010 & 57305.62  & PC    & 1.0  & $0.060\pm0.008$ \\
33770011 & 57306.67  & PC    & 1.0 & $<$ $0.008$ \\
33770012 & 57307.68  & PC    & 1.0 & $<$ $0.013$ \\
33770013 & 57308.63  & PC   & 1.0  & $<$ $0.008$ \\
33770014 & 57309.27  & PC   & 1.0  & $<$ $0.015$  \\
33770015 & 57310.54  & PC   & 0.8  & $<$ $0.016$ \\
33770016 & 57311.43  & PC   & 1.0  & $0.005\pm0.003$ \\
33770017 & 57660.69 & WT    & 1.0  & $26.3\pm0.8$ \\
33770018 & 57661.35 & WT    & 0.8 & $19.0\pm0.1$ \\
33770019 & 57662.75 & WT    & 0.4  & $4.7\pm0.6$ \\
33770020 & 57663.08 & WT    & 0.036   & $3.5\pm0.4$ \\
33770020 & 57663.35 & PC    &  1.7      & $1.3\pm0.05$ \\ 
33770021 & 57664.18 & PC    &  1.7  & $0.109\pm0.008$  \\
33770022 & 57665.61 & PC    & 1.4   & $<$ $0.008$  \\
33770023 & 57667.52 & PC    &  0.6   & $<$ $0.016$  \\
33770024 & 57668.09 & PC    & 2.0   & $0.002\pm0.001$  \\
33770025 & 57669.76 &  PC   & 1.7   & $<$ $0.007$  \\

\hline
\hline
\end{tabular}
\label{Swift}
\\

\end{table*}

\begin{figure*}
\centering
\begin{minipage}{0.45\textwidth}
\begin{flushleft}
\includegraphics[height=4.in,width=8.cm,angle=0,keepaspectratio]{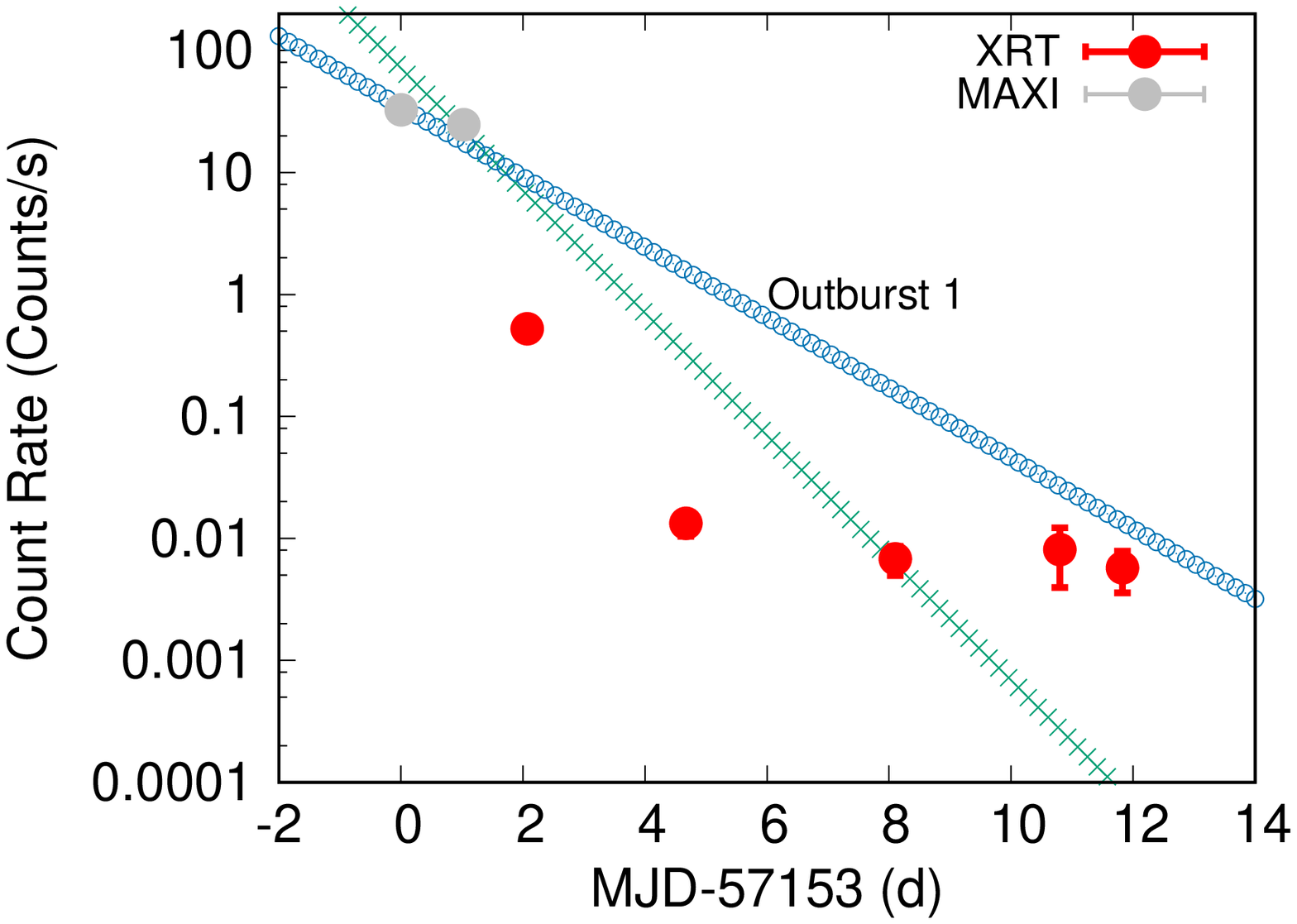}
\includegraphics[height=4.in,width=8.cm,angle=0,keepaspectratio]{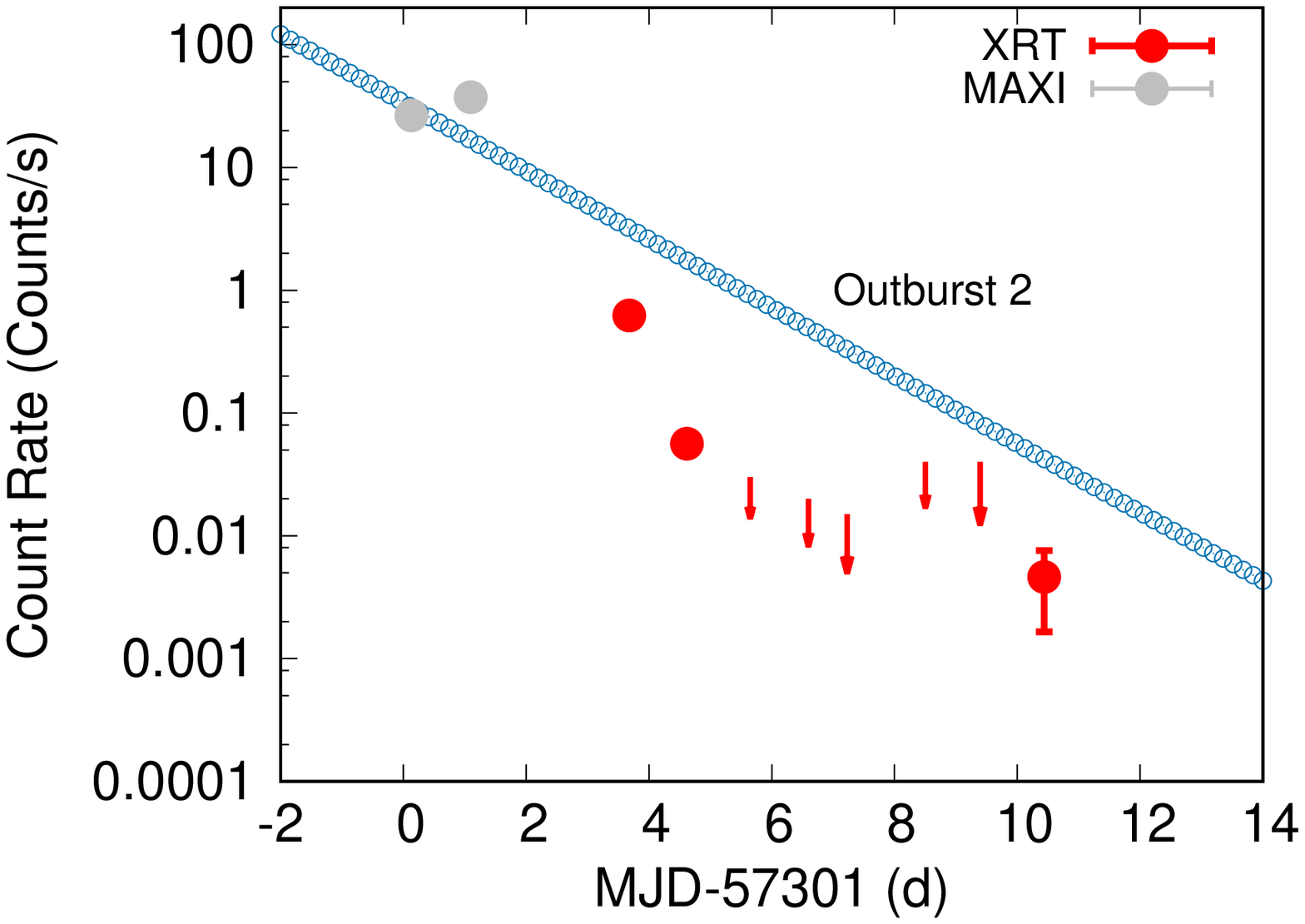}
\end{flushleft}
\end{minipage}
\hspace{0.035\linewidth}
\begin{minipage}{0.45\textwidth}
\begin{flushright}
 \includegraphics[height=4.in,width=8.cm,angle=0,keepaspectratio,trim=0cm 0cm 0cm 0.5cm]{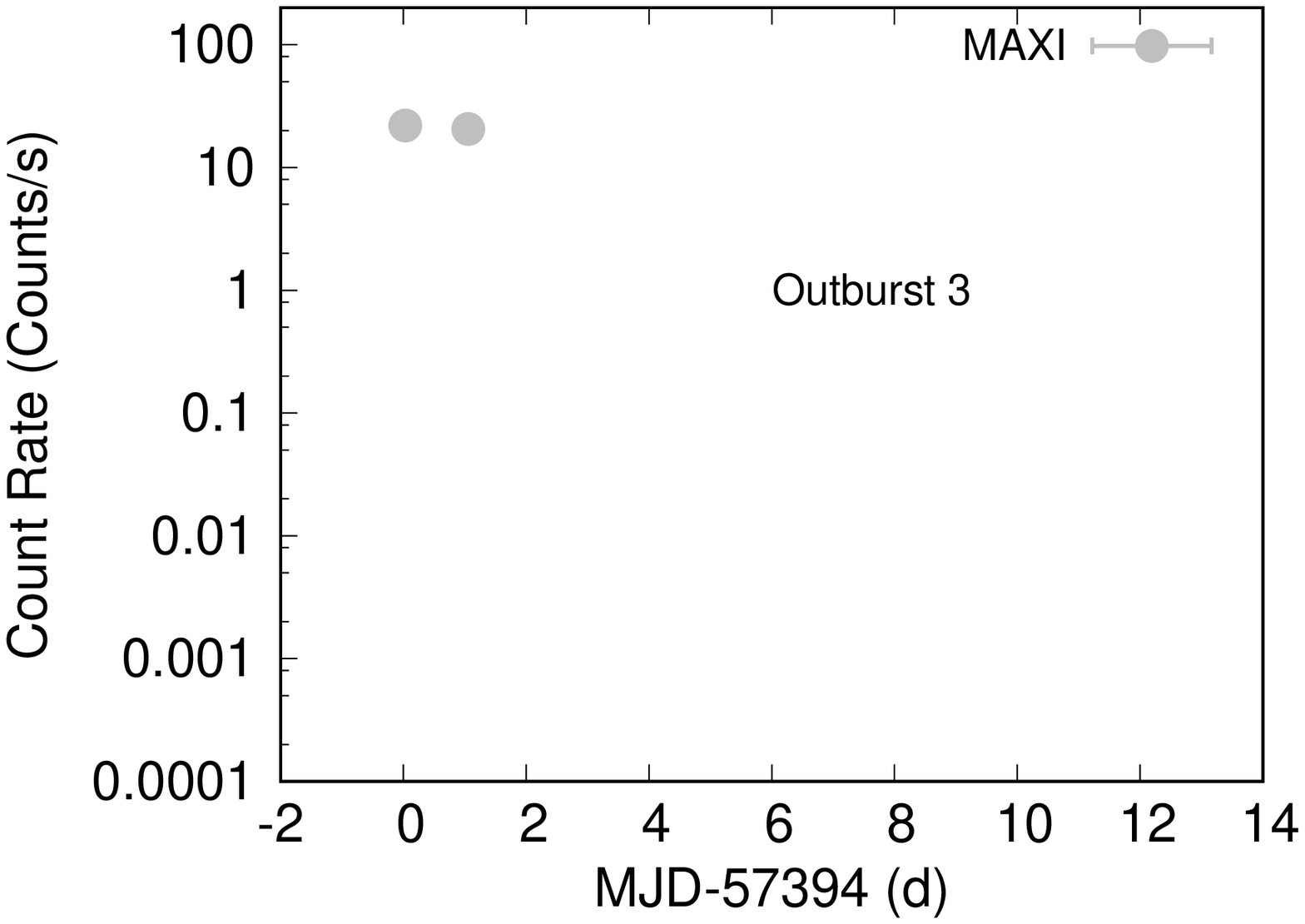}
\includegraphics[height=10.5in,width=8.cm,angle=0,keepaspectratio,trim=0cm 0.55cm 0cm 0cm]{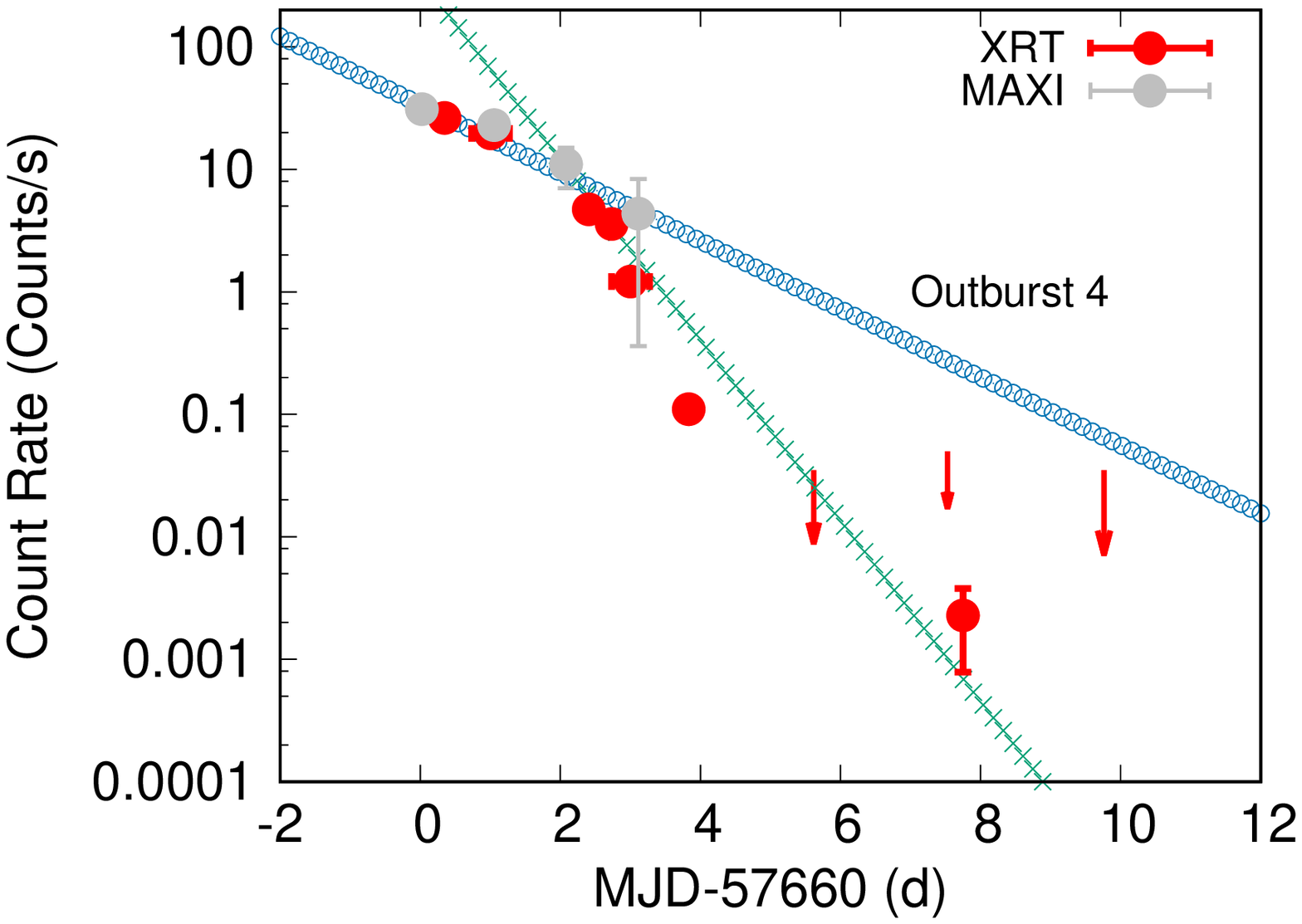}
\end{flushright}
\end{minipage}
\caption{The four
  outbursts observed in MAXI~J1957+032.  The plot at the top-left
  corresponds to the first outburst which began on
  2015-05-11~(MJD~57153), the bottom-left shows the second
  outburst~(beginning on 2015-10-06,~MJD~57301), the top-right is for the third
  outburst~(started on 2016-01-07,~MJD~57394) which was only observed with
  \emph{MAXI}, and the fourth outburst which starting on
  2016-09-29~(MJD~57660) is shown in the bottom-right panel.  The
  arrows correspond to the upper limits of the
  detection with \textsc{XRT}.~We show \textsc{XRT} count rate in log scale.~The values of flux measured with \emph{MAXI} 
  were converted to \textsc{XRT} count rates and are shown in grey.~For the \emph{MAXI} data we have included only
  those data points which show significant detection (above 4~${\sigma}$).~We also plot exponential decays where 
 the crossed line~(green) shows {\bf the fast decay while the empty circles~(blue) is for the slow decay~(see Section~3.1 for e-folding timescales).}}
\label{Outbursts}
\end{figure*}


\begin{table*}
\def\arraystretch{1.1}%
\caption{Best fit parameters of MAXI~J1957+032 obtained using the \emph{Swift}-\textsc{XRT} data.
Spectral fitting was performed using  W-statistics.
}
\label{my-label} 
\begin{tabular}{cc |cccc}
 \hline
       &              &          &       &     \\ [0.5ex]       
 
Outburst  & Obs~ID & N$_H$ (free) & $\Gamma$ & $N_{PL}$ & Unabsorbed flux  \\ [0.5ex] 

              &        &              &          &            &  (0.5-10~keV)      \\ [0.5ex]   
             
              &        & (10${^2}{^1}$\rm{atoms cm$^{-2}$}) &  & ($\rm{photons~cm^{-2}~s^{-1}~keV^{-1}}$) at 1~keV & ($10^{-10} \rm{ergs~cm^{-2}~s^{-1}}$) \\
\hline

Outburst~1 & 00033770001 & $3.6\pm0.7$ & $2.5\pm0.2$ & $0.007\pm0.001$ & $0.15\pm0.01$ \\ [0.5ex]

\hline
Outburst~2 & 00033770009 & $3.7\pm0.8$ & $2.8\pm0.3$ & $0.010\pm0.003$ & $0.33\pm0.02$  \\ [0.5ex] 

& 00033770010 & $4.9_{-2.7}^{+3.9}$ & $4.87_{-1.5}^{+2.2}$ & $0.002\pm0.001$ & $0.07\pm0.02$  \\ [0.5ex] 

\hline
& 00033770017 & $1.1\pm0.2$ & $1.78\pm0.04$ & $0.175\pm0.008$ & $10.1\pm0.1$  \\ [0.5ex] 
& 00033770018 & $1.7\pm0.2$ & $2.01\pm0.05$ & $0.146\pm0.008$ & $6.9\pm0.1$ \\ [0.5ex] 

Outburst~4 & 00033770019 & $4.7\pm2.0$ & $2.2\pm0.2$ & $0.04\pm0.01$ & $1.7\pm0.1$ \\ [0.5ex] 

& 00033770020  & $3.1\pm0.7$ & $2.49\pm0.22$ & $0.017\pm0.003$ & $0.61\pm0.03$  \\ [0.5ex] 
& 00033770021 & $2.0\pm1.0$ & $2.6\pm0.4$ & $0.0013\pm0.0004$ & $0.045\pm0.005$\\ [0.5ex]

%
%
\hline

\end{tabular}

{\bf{Notes}}: 
Errors quoted are for the 90 $\%$ confidence range.~The energy range used is 0.5-10~keV. 
$N_{PL}$ is the Normalisation of power law~(PL) \\

           \label{Best-fit1}  
\end{table*}


The \textsc{XRT} observations were performed in photon counting (PC) and
windowed timing (WT) modes depending on the brightness of the source.  We have used
the online tools provided by the UK Swift Science Data
Centre\footnote{http://www.swift.ac.uk/} \citep{Evans09} to obtain the
spectrum of each observation.~Some of the spectra showed low number of
photons, therefore, did not allow
us to use $\chi^{2}$ statistics.~To maintain the homogeneity in our analysis, we 
grouped the obtained spectra using the ftools task \textsc{grppha}~(\textsc{HEASOFT} v6.19) to have at least one count per bin.
Spectra were fitted using
the \textsc{XSPEC 12.9.1} \citep{Arnaud96}.~We have used W-statistics which is background subtracted Cash
statistics \citep{Wachter79}. \\

To obtain insight in the nature of the accretor in our target
we compared our results with that of \citet{Wijnands15}.
We fitted an absorbed power-law to our spectra.~\textsc{tbabs} was used to model
the hydrogen column density~($N_\mathrm{H}$) using \textsc{WILM} abundances
\citep{Wilms00}.~The
values of $N_\mathrm{H}$ used are discussed in the next section.~For the observations made in PC mode, 
we fitted the spectra between 0.5 and 10~keV. The WT spectra are fitted
in the range of 0.7-10~keV as there exist low energy spectral residuals
below 0.7~keV in the WT mode spectra\footnote{e.g., see: http://www.swift.ac.uk/analysis/xrt/}.
All the fluxes reported are the unabsorbed fluxes and we have used the convolution model
{\textquoteleft}\textsc{cflux}' to measure the fluxes in the 0.5--10~keV range. 
In the Appendix, we show that an absorbed power-law provides a good fit
to the \textsc{XRT} observations of MAXI~J1957+032 where we have used only spectra of sufficient quality
that allowed us to use $\chi^{2}$ statistics to determine
the goodness of the fits.
Figure A1 in the Appendix shows the best spectral fit obtained with the brightest observation
made with \textsc{XRT}.~Table~A1 gives the fit parameters obtained using an absorbed power-law. \\

The Monitor of All Sky X-ray image (\emph{MAXI}) is an all sky monitor
\citep{Matsuoka09} which has two instruments on board: the Solid state
Slit Camera (\textsc{SSC}), operating between 0.7-7 keV
\citep{Tomida11} and \textsc{GSC}, operating in the
energy range 2-20 keV \citep{Mihara11}.~\textsc{GSC} has a wider field-of-view
and much larger collecting area than \textsc{SSC}.
MAXI~J1957+032 was detected
with the \emph{MAXI}-\textsc{GSC} during all the four outbursts. 
We have used \emph{MAXI}-\textsc{GSC} light curves extracted
in the 2-10~keV energy band using the \emph{MAXI} on-demand
data processing{\footnote{http://maxi.riken.jp/mxondem/}}~\citep{Nakahira10}. \\


\begin{figure}
\captionsetup[figure]{font=small,skip=0pt}
\includegraphics[height=4.in,width=9.5cm,angle=0,keepaspectratio,trim=0cm 8cm -2cm 0cm]{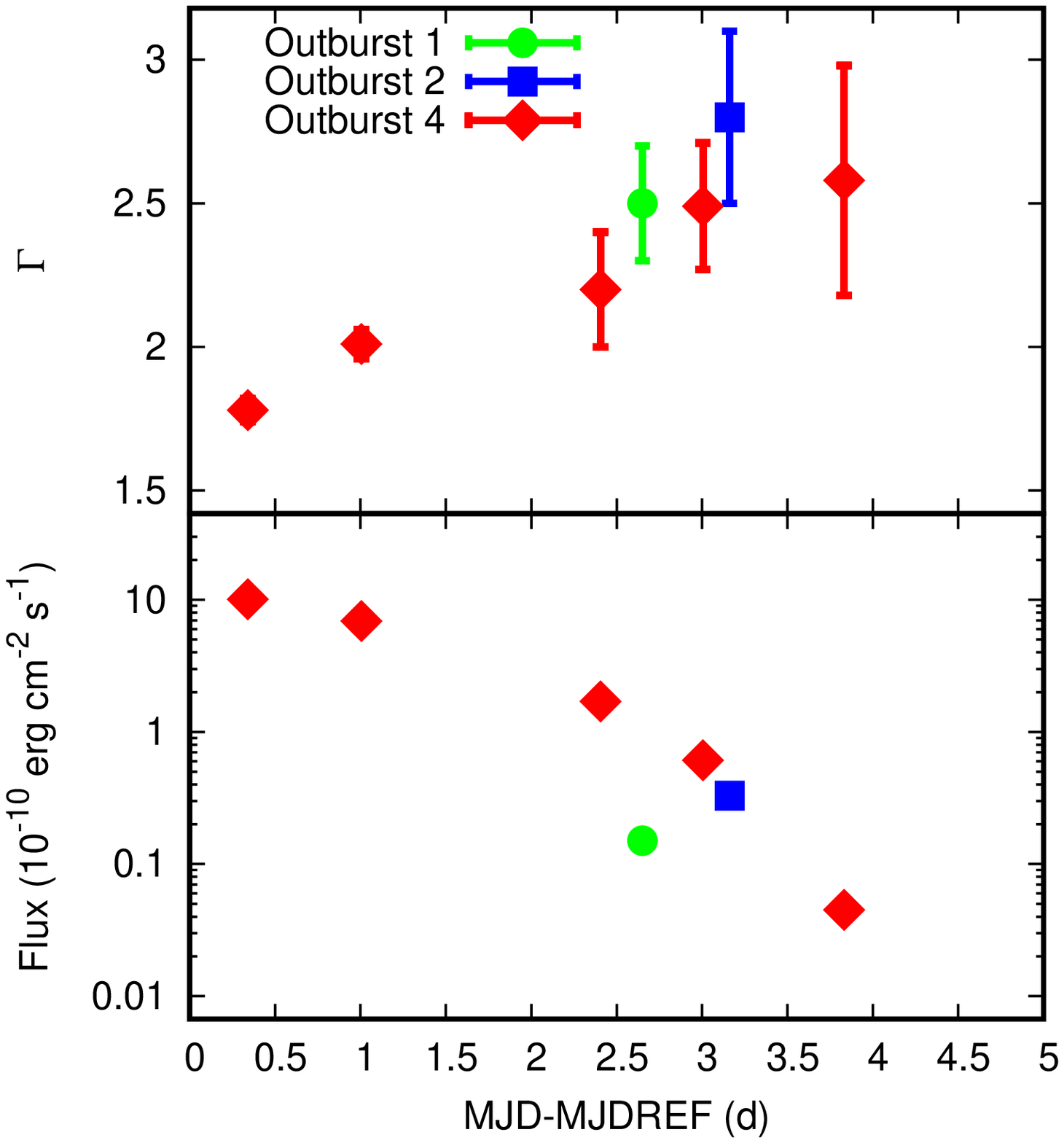}
\caption{Variation of the index, and the unabsorbed
  flux~(0.5--10 keV) of the power-law model during the three outbursts observed with \emph{Swift}-\textsc{XRT} in 2015 and 2016.
  The power-law index was estimated keeping $N_\mathrm{H}$ as a free parameter.
  The spectrum obtained during the \emph{Swift}-\textsc{XRT} observation made on MJD~57305.62~(outburst~2)
  had low statistics and consequently the errors
  on the obtained photon index were very large~($\Gamma$ has an error
  greater than 0.5).~Therefore, this point was not plotted.}
 \label{Spec-var}
\end{figure}


\section{Results}

\subsection{Light Curves}
MAXI~J1957+032 underwent four outbursts that started on 2015-05-11~\citep[outburst~1;~ATel 7504;][]{Negoro15},
2015-10-06~\citep[outburst~2; ATel 8143;][]{Sugimoto15},
2016-01-07~\citep[outburst~3;~ATel 8529;][]{Tanaka16} and 2016-09-29~\citep[outburst~4;~ATel 9565;][]{Negoro16}.
The four outbursts of MAXI~J1957+032 were observed with \emph{MAXI} but only three of these 
were monitored with \emph{Swift}.~The third
outburst (outburst~3) was not covered with \emph{Swift}-\textsc{XRT} due
to the Sun angle constraint.
In Figure~\ref{Outbursts} we show combined light curves of MAXI~J1957+032, obtained using \emph{MAXI}-\textsc{GSC}
and \emph{Swift}-\textsc{XRT} data, however, for the third outburst we show only the \emph{MAXI}
light curve.~We have used MJD~57153, MJD~57394,~MJD~57301,~MJD~57660 as
the reference time for outburst~1, outburst~2, outburst~3, outburst~4 respectively.
We have used \textsc{WebPIMMS~HEASARC} tool to convert \emph{MAXI}-\textsc{GSC} count rates obtained in the 2-10~keV energy band 
to \emph{Swift}-\textsc{XRT} count rates
in the 0.3-10~keV energy band.~The best-fit values of $N_\mathrm{H}$ and $\Gamma$ obtained using 
the spectral fitting of the brighest observation
made with \textsc{XRT} during each outburst were used for this conversion (refer Table~2).~However, for the third outburst, 
we assumed the
value of 
$N_\mathrm{H}$ and $\Gamma$ as observed during its outburst~4. \\

%
~Figure~\ref{Outbursts} shows the evolution of four outbursts observed in
MAXI~J1957+032 and it can be seen that all the four outbursts were equally bright.
However, only during the fourth outburst~(outburst~4) \textsc{XRT} observations were made close to the peak
of the outburst.
The peak count rate during the fourth
outburst observed with \emph{Swift}-\textsc{XRT} is $\sim$~26~$\mathrm{count~s^{-1}}$ 
which is almost a factor of 50 higher than observed during the other two outbursts.
These outbursts do not last more than a few days, and in all cases \emph{Swift}-\textsc{XRT} observations 
sample (part of) the outburst decay \citep{Ravi17}.
We fitted an exponential decay function to the decay
to obtain the e-folding time of these outbursts.
The decay time scale obtained for outburst~2 is $0.64\pm0.08$. Here, the exponential decay function
was fitted starting from the peak of the outburst. 
Outburst~1 and 4 indicated the presence of two e-folding times. 
Therefore, we fitted these two outburst curves using an exponential decay function in two different time ranges.
This allowed us to determine slow and fast declines during these outbursts
(see Figure~\ref{Outbursts}).
The e-folding timescales of outburst~1 are $1.16\pm0.22$, and $0.66\pm0.11$ days corresponding to the slow and fast declines respectively.
For the outburst~4, we obtained $1.7\pm0.4$, and $0.64\pm0.08$ days to be the slow and fast decay timescales respectively

These obtained decay timescales are consistent with that
reported by \citet{Mata17}.
For the third outburst~(outburst~3) it is difficult to determine the decay time scale using the \textsc{GSC} data.
The errors quoted on the decay timescales are within $90~{\%}$ confidence range.

\subsection{Absorption Column Density~($N_\mathrm{H}$)}

From the spectral fitting we found that,~when it was left free in the fits,
the value of $N_\mathrm{H}$ showed a large variation throughout the brightest outburst observed in 2016.
To investigate, if the higher value of column density is
due to the requirement of a thermal X-ray component,~we tried fitting the spectra with a model composed of 
a soft thermal component~(\textit{bbodyrad}) and a power law component. 
We noticed,~however, that thermal component was not required in 
all these spectra.~During the brighest outburst of MAXI~J1957+032, only one of the
observations showed the presence of a soft component (for details see Section~4).
Thus, a higher value of $N_\mathrm{H}$ might indicate the increase in the intrinsic absorption
with an decreasing X-ray luminosity.~Another possibility, could be that at higher
X-ray luminosities the source is far away from a true power law shape. This was also suggested 
by \citet{Cherepashchuk15} when the authors found that the simple extrapolation of the \textsc{XRT} spectrum
of MAXI~J1957+032 to the 20--60~keV
energy band
resulted into five times lower value of X-ray flux compared to that measured with \emph{INTEGRAL}.

\subsection{Spectral Evolution during Outbursts of MAXI~J1957+032}

In Figure~\ref{Spec-var} we show the spectral evolution of
MAXI~J1957+032 during the three outbursts as observed with \emph{Swift}-\textsc{XRT}.
The two panels (from top to bottom) are for the power-law index,
and the unabsorbed flux in the 0.5--10~keV band respectively.
We found that while $\Gamma$ generally increases with time, and the
0.5--10~keV absorbed flux decreased; clearly
there is an anti-correlation between power-law index and observed
flux.
During the brightest outburst of MAXI~J1957+032 in 2016, the values of
power-law index~($\Gamma$) increased from $\sim$~1.8 to $\sim$~2.5
when the fluxes decreased (outburst~4, Figure~\ref{Outbursts}).
From the spectral analysis of MAXI~J1957+032, we found
that $\Gamma$ reaches the value close to 2.5 during its three outbursts
observed with \emph{Swift}-\textsc{XRT} \citep[see also][]{Mata17, Ravi17, Kennea17}.

%

\subsection{Photon Index versus luminosity}

We converted the unabsorbed fluxes in the 0.5-10~keV band
(given in Table~\ref{Best-fit1}) to X-ray luminosities. The distance
to MAXI~J1957+032 is not known; therefore, below we discuss different
interpretations of our data based on a wide range of
distances.
Figure~\ref{Gamma-Lumin} shows $\Gamma$ as a function of X-ray
luminosity in the 0.5-10~keV band.  This plot includes all the data
points used by \citet{Wijnands15} and also the values we have obtained for
MAXI~J1957+032.~The black and the grey points of the
Figure~\ref{Gamma-Lumin} corresponds to BH and the NS
binary systems, respectively.  The red,~blue, and green points are of
MAXI~J1957+032, obtained using \emph{Swift}-\textsc{XRT} data while the
orange point corresponds to the photon index obtained using the \emph{Chandra} observation of MAXI~J1957+032
during outburst~4 \citep{Chakrabarty16}.

\section{Discussions}
In our work, we have used \emph{Swift}-\textsc{XRT} observations during 
three of four outbursts of MAXI~J1957+032.
MAXI~J1957+032 is believed to be a VFXT and the nature of its compact 
object is still not established.
We observe that the light curves of MAXI~J1957+032 during its outbursts show a very short 
exponential decay timescales of less than a day.
Based on these characteristics like short outbursts and short recurrence time of outbursts,
MAXI~J1957+032 {\bf was} proposed to be an AMXP similar to NGC~6440~X--2 by \citet{Mata17}.
However, there are several VFXTs in the Galactic center which are not AMXPs but show
short duration outbursts \citep[see e.g.,][]{Degenaar15} and there are also several
bright NS X-ray transients (not AMXPs) that show outbursts which last only for a few days
with a peak luminosity of $<$$10^{36}~\mathrm{ergs~s^{-1}}$ for example,~GRS 1741-2853 \citep{Degenaar10}, XTE~J1701--407
\citep{Fridriksson11}, Aql~X--1 \citep{Coti13}, SAX J1750.8-2900 \citep{Wijnands13}. \\

Outbursts are often believed to be due to accretion disk instabilities.
According to standard accretion disk instability 
models e.g., \citet{Lasota01} one would expect brighter outbursts to be longer. 
If MAXI~J1957+032 is a short period binary system then one would expect a long interval between
outbursts \citep[also see][for details]{Heinke10}.~MAXI~J1957+032 showed 
an increased activity in a year and half starting from June 2015.~It showed outbursts 
once every few-hundred days.~\emph{MAXI}/\textsc{GSC} nova alert system~\citep{Negoro-PASJ16} has triggered on
four outbursts from MAXI~J1957+032. To search for other
brightening episodes of MAXI~J1957+032, we checked the publicly available \emph{MAXI}-\textsc{GSC}
light curves which include data before its first outburst~(starting from August 2009), however, we did not find
any flaring event prior to the its first X-ray outburst~(outburst~1). 
Another possible cause of these outbursts is mass-transfer variations from the donor
similar to that suggested for NGC~6440~X--2 \citep{Heinke10}.
If MAXI~J1957+032 is a triple star system as proposed by \citet{Ravi17}
one might expect to observe change in the orbital parameters, induced 
by the distant companion.~However, this would need monitoring of future outbursts
with more sensitive instruments like \emph{XMM-Newton, Chandra}. \\

%

\begin{figure*}
\centering
 \includegraphics[height=2.8in,width=10.cm,angle=0,keepaspectratio]{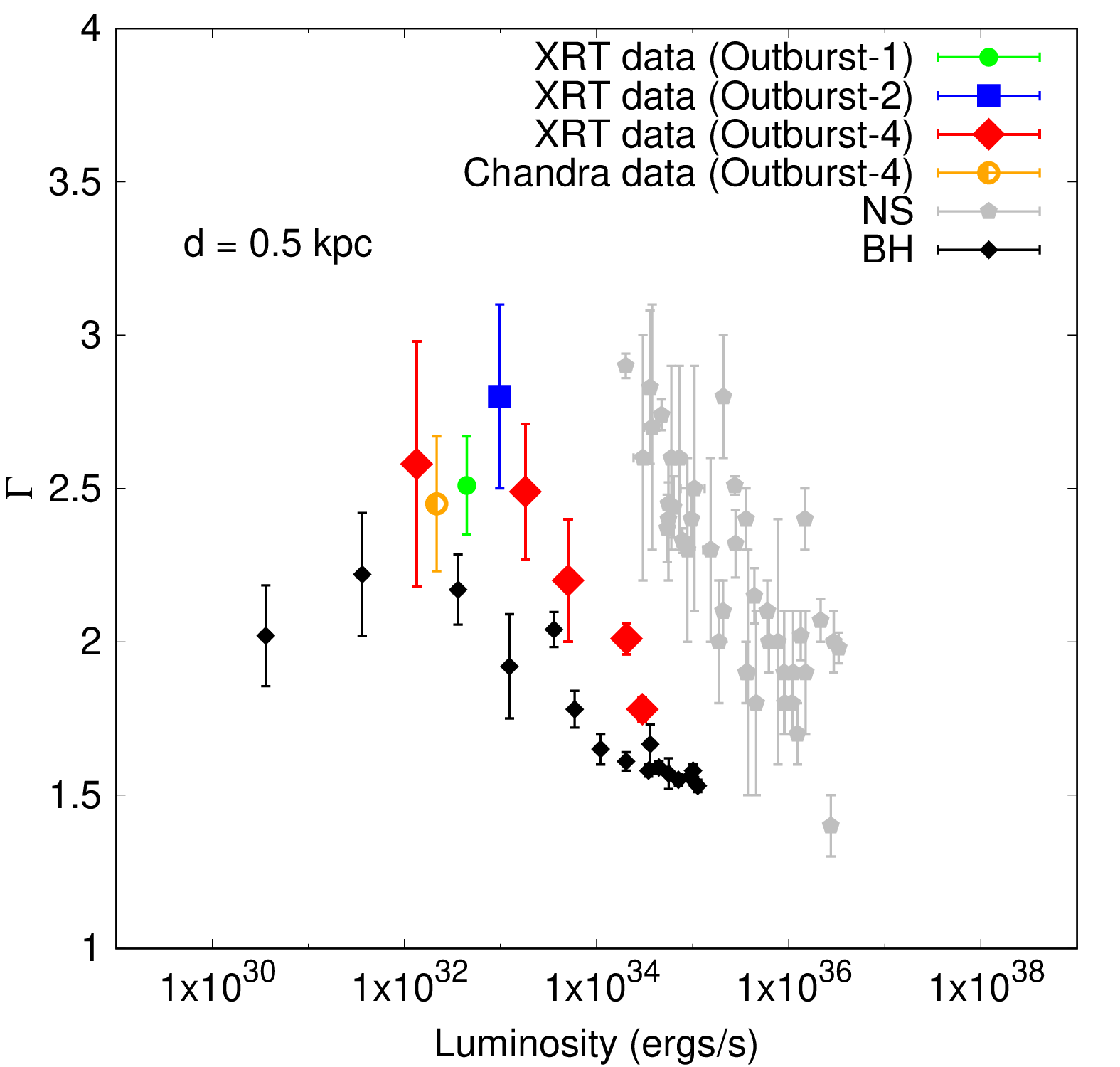}
 \includegraphics[height=2.8in,width=10.cm,angle=0,keepaspectratio]{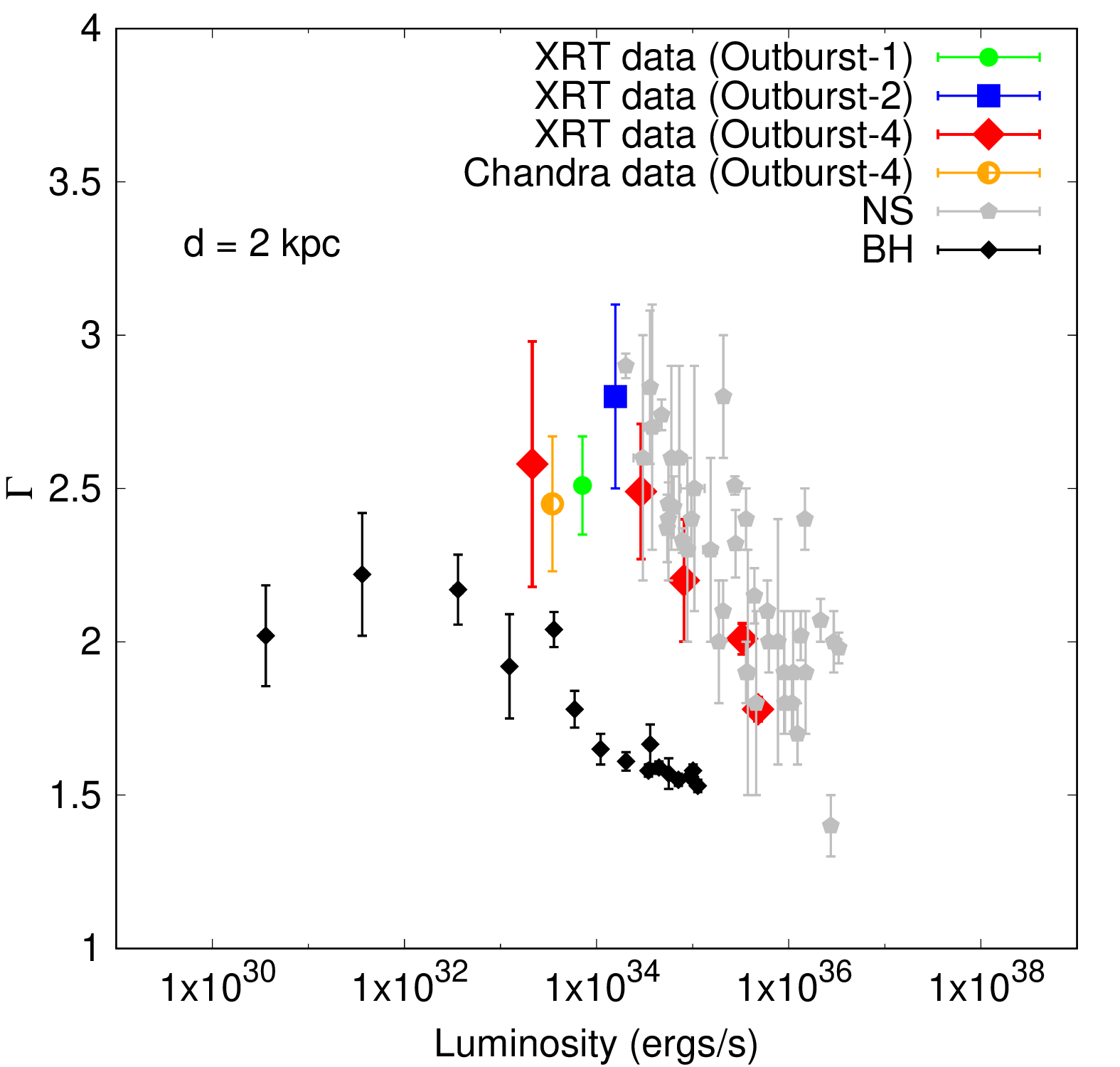}
 \includegraphics[height=2.8in,width=10.cm,angle=0,keepaspectratio]{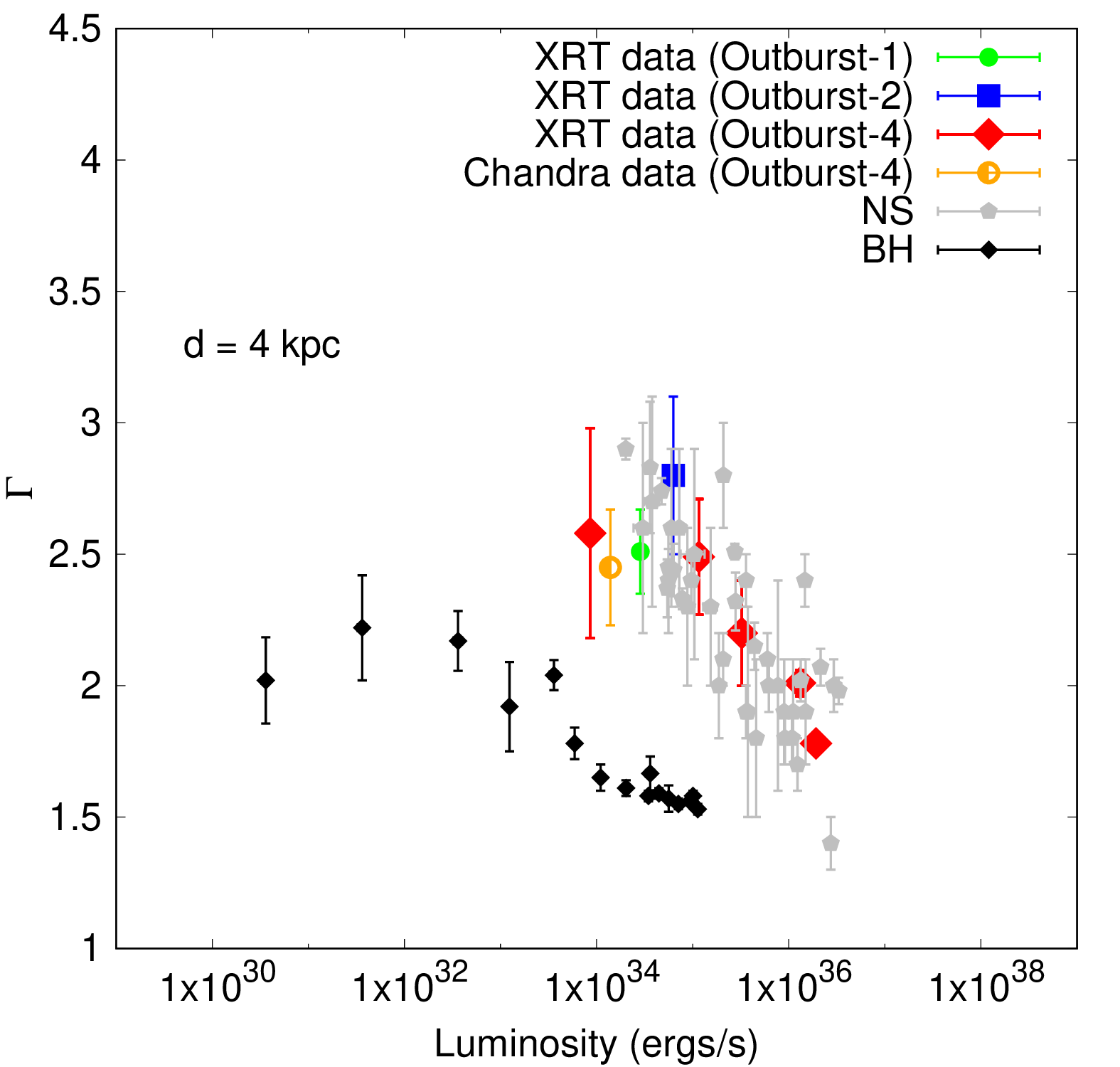}
 \includegraphics[height=2.8in,width=10.cm,angle=0,keepaspectratio]{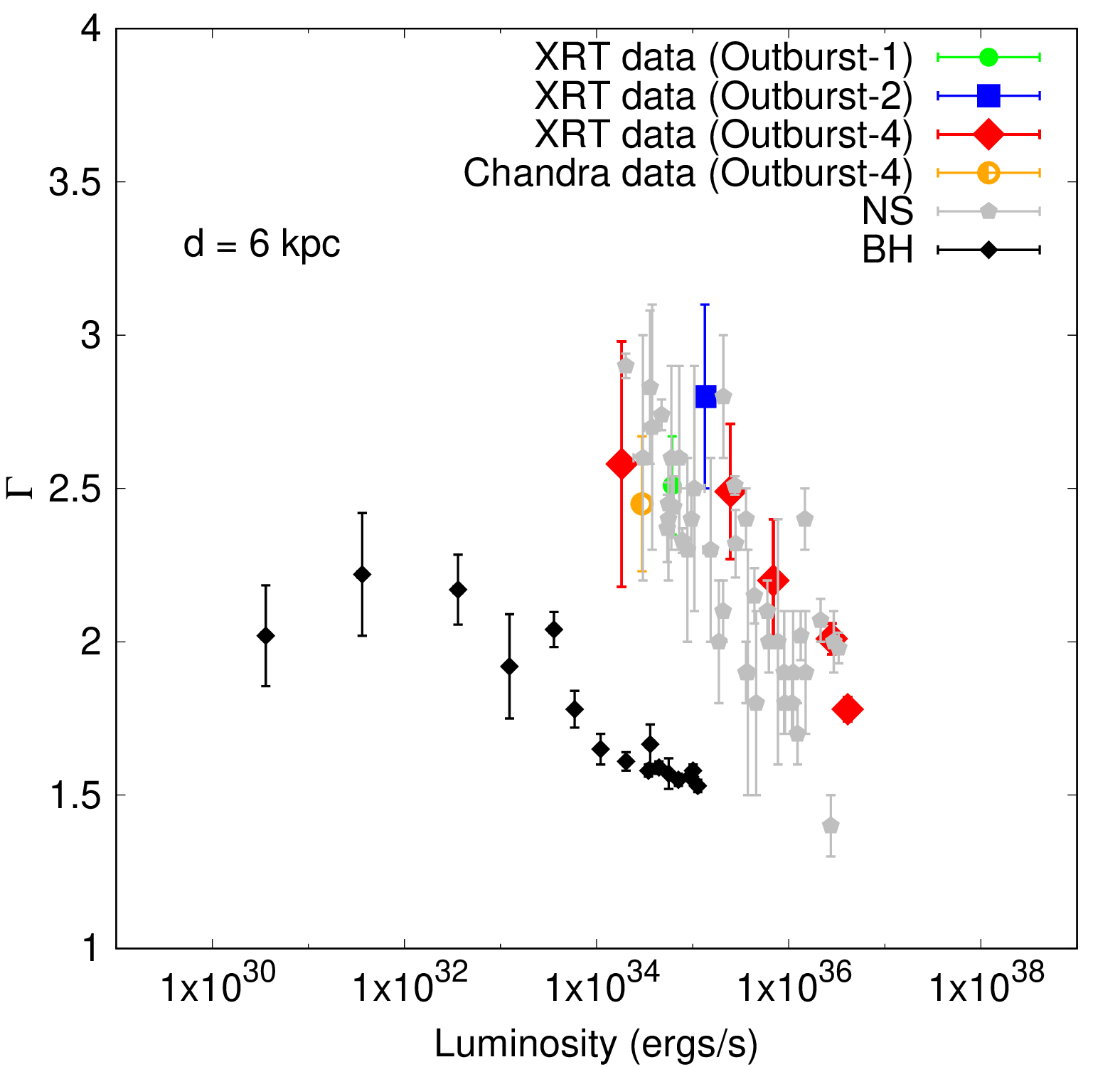}
 \includegraphics[height=2.8in,width=10.cm,angle=0,keepaspectratio]{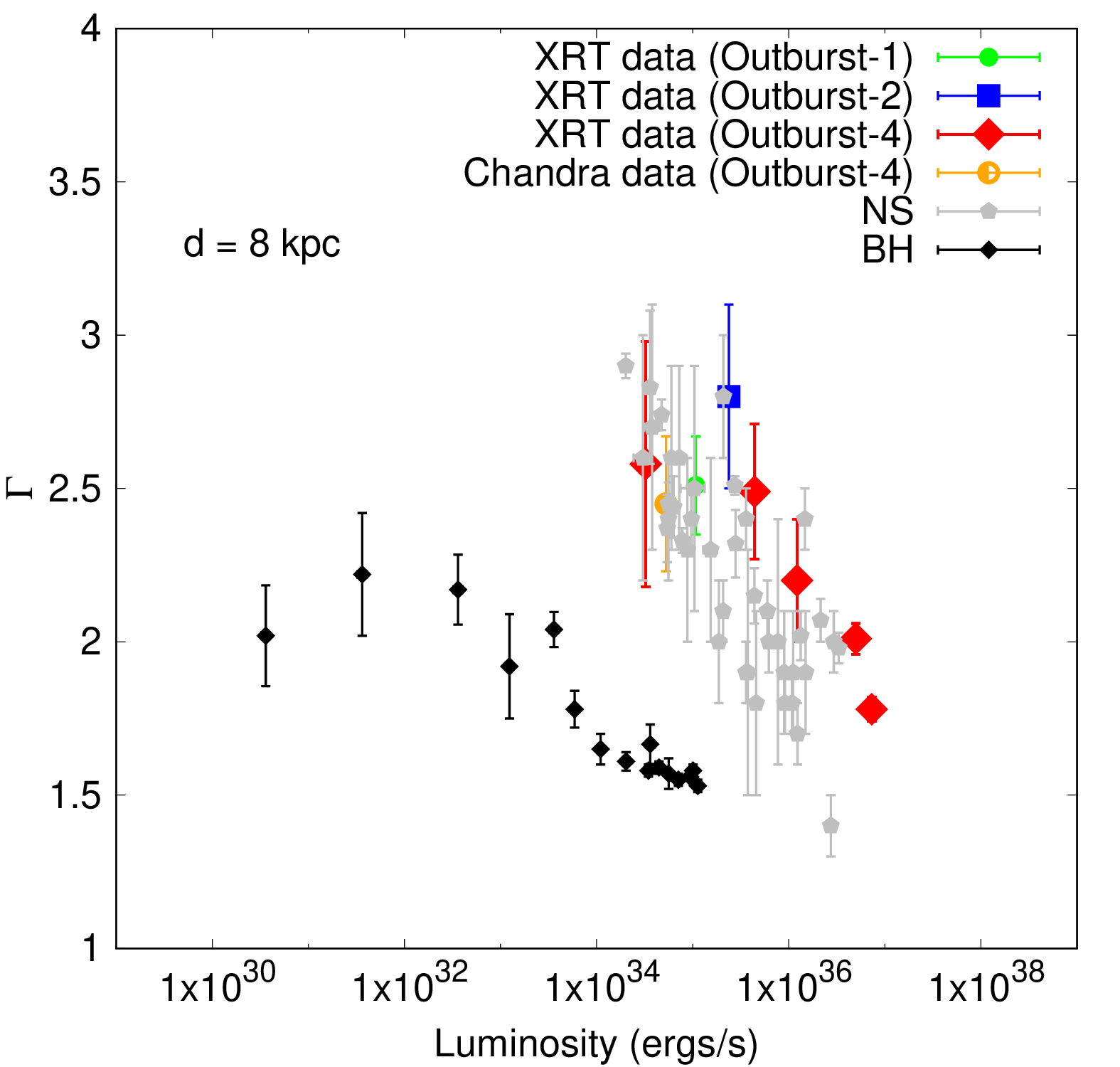}
 \includegraphics[height=2.8in,width=10.cm,angle=0,keepaspectratio]{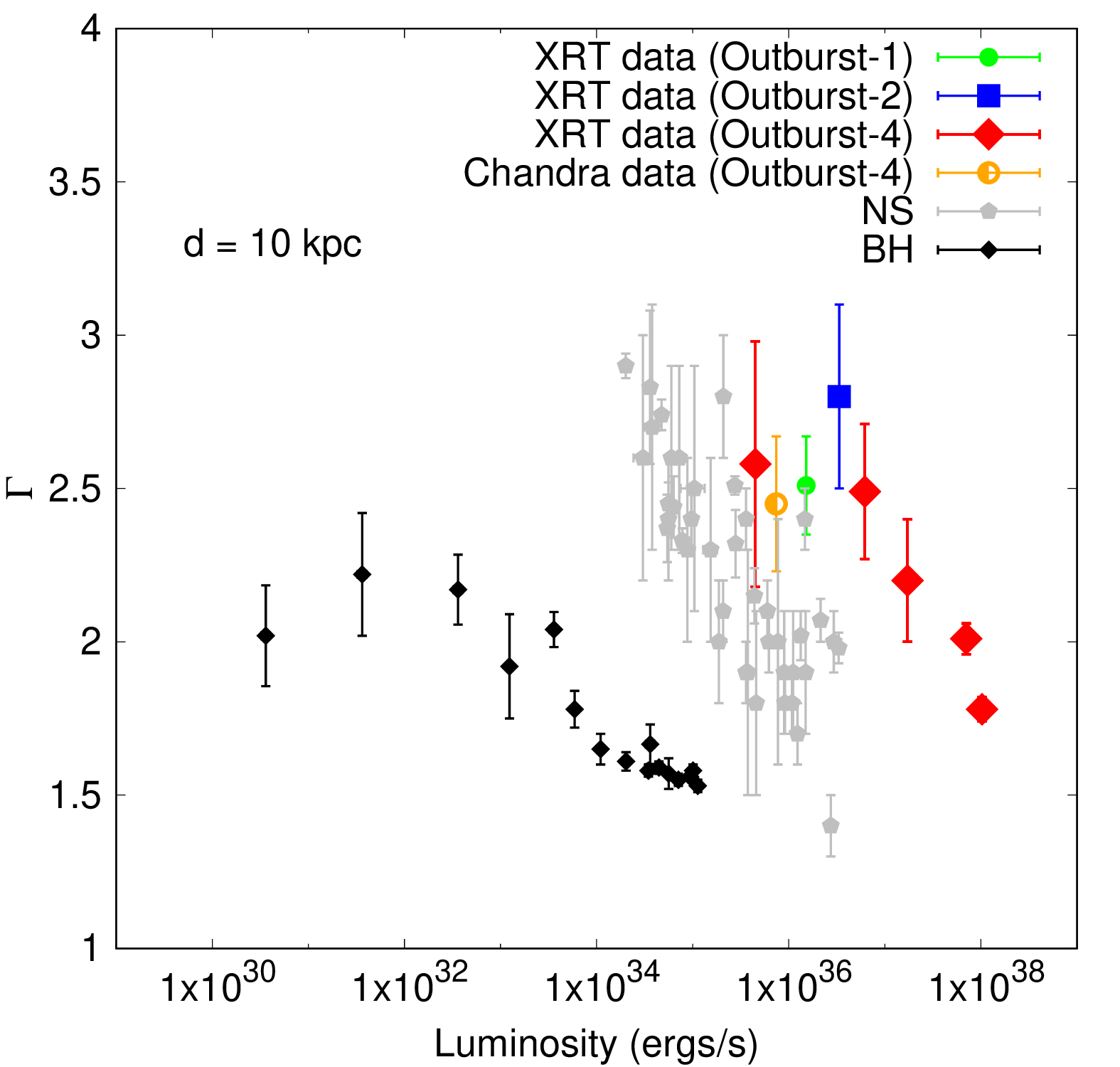}
\caption{The photon index versus the 0.5-10~keV X-ray luminosity for MAXI~J1957+032, including the data presented by \citet{Wijnands15}
for NS and BH binaries.}
\label{Gamma-Lumin}
\end{figure*}

We have studied the spectra (using an absorbed power-law model) and we observed an
anti-correlation between $\Gamma$ and the total flux similar to what has been observed
in the spectra of many other LMXBs \citep[see
  e.g.,][]{Armas11,Armas13,Reynolds14}. The spectra become softer
as the source luminosity decreases \citep[see also][]{Kennea2015,Kennea-2016}.
  After comparing the spectra of several LMXBs (NSs and BHs) \citet{Wijnands15}
reported that NS binary systems tend to be much softer
compared to the BH systems at low luminosities.  These authors
found that in the case of NS X-ray binaries at low
luminosities~($10^{34}-10^{35}$ $\mathrm{ergs~s^{-1}}$) the value of $\Gamma$ can be
as high as 3.~We explore different possible scenarios based on the assumption of
different values of source distance of MAXI~J1957+032.
Figure~\ref{Gamma-Lumin} shows how different distances affect our conclusions
which we discuss in more detail here.~If the distances of MAXI~J1957+032 is as low as 0.5~kpc, our data 
lay closer to the the track of BHs rather than a NS.
   In addition, at the lowest X-ray luminosities, 
   the two data points of MAXI~J1957+032 are significantly softer than the BH data points.
   We also note that although the errors on $\Gamma$ estimated for rest of the data 
   points of MAXI~J1957+032 are large,~the trend clearly shows higher values of $\Gamma$ 
   than seen for the BH data points.~Moreover, we note that \citet{Ravi17}
   proposed the source distance to be $\sim$~5~kpc using the R-band magnitude for the MAXI~J1957+032
   counterpart.~The authors also suggested that the V-band extinction and measured hydrogen column density
   are consistent with the proposed source distance.
   

For the source distance between 2 to 8 kpc, our data
better follows the track of the NS LMXBs.  At 4~kpc,
MAXI~J1957+032 falls in the regime of VFXTs that have maximum 
X-ray luminosities in outburst between $10^{34}-10^{36}~erg~s^{-1}$ \citep{Wijnands06}. 
In that case, at around
$10^{35}$ $\mathrm{ergs~s^{-1}}$ thermal emission from the neutron star surface might
become visible and the spectrum should then be described with an absorbed
\textit{bbodyrad+power-law} model \citep[see discussions in][]{Wijnands15}.
Therefore, we re-fitted the spectra by adding
 \textit{bbodyrad} component. 
 We found that on adding the \textit{bbodyrad} component, the
 the value of $\chi^2$ decreased from 50 to 43 for the observation with ID~00033770001
 and for the observation with ID~00033770020, the value of $\chi^2$  decreased from 17 to 13
 for 2 degrees of freedom~(dof) less.~Thus, the 
 spectra of two observations (ID~00033770001 \& 00033770020;
 MJDs: 57156.03 and 57663.35 respectively)
showed a potential contribution of thermal emission in the form of blackbody
The best-fit parameters are given in Table~\ref{Best-fit3}.~However, 
we would like to mention that some of the fit parameters are not well constrained
owing to the limited statistics.
These observations corresponds to the green and fourth red point from
the left (between $10^{34}$ and $10^{35}$ $\mathrm{ergs~s^{-1}}$) of
Figure~\ref{Gamma-Lumin} at d=4kpc.
We observe that $\Gamma$ is harder when a blackbody component is included.

This has also been seen in other VFXTs such as XTE J1709--267,
IGR~J17494--3030 which have both been proposed to be NSs
based on their X-ray spectral properties \citep[][]{Degenaar13a,Armas-Padilla13} \citep[see also][for a discussion]{Wijnands15}.
Moreover, the confirmed NS VFXT and AMXP IGR~J17062--6143 also shows a thermal component in its spectrum \citep{Degenaar17}.
The obtained $kT$ values and power law that we obtained for MAXI~J1957+032 are consistent with studies of 
these systems.
We note that in the black hole LMXB and VFXT, namely,~Swift~J1357.2-0933 the high-quality \emph{XMM-Newton}
data also required a blackbody component but the contribution of the
soft~(thermal) component to the total flux is less than 10$\%$ \citep{Armas13},~while we
find that it contributes about 30-40~$\%$ of the total flux observed for MAXI~J1957+032. \\

We also note that \citet{Armas-Padilla13,Shidatsu17,Degenaar17} suggested the 
possible origin of the observed thermal emission in the X-ray spectra to be neutron
star surface or the accretion disc.~From
Table~\ref{Best-fit3}, we observe that the obtained values of blackbody radius~(2-3~kpc)
are very similar to that have been found in other systems that are either proposed to be NS or
are confirmed NS systems.~Thus,~the origin of the thermal emission can be a part of the neutron star. \\

There are systems even as far as 20-50 kpc \citep[e.g., the BHC GS~1354-64 at a distance of about 25 kpc, NS~MAXI~J0556--332
  at a distance $\sim$46~kpc see;][]{Casares04,Homan14}.
However, we note that a very recent distance measurements performed by \citet{Gandhi18}
using data obtained with \emph{Gaia} suggests that GS~1354-64 is a nearby system~($\sim$~0.6~kpc).
If we assume that MAXI~J1957+032 is at about $8-10$ kpc, then the
highest flux levels we measured would correspond to a source
accreting at $\sim10^{38}$ $\mathrm{ergs~s^{-1}}$. The softening we observed could
correspond to the softening that it is sometimes seen in some BH and
NS systems at high luminosities \citep[see, e.g., ][and references
therein; however note that these works generally use
multi-component models, making it difficult to make precise
comparisons]{Remillard06, Lin07, Soleri13, Fridriksson15}. 
We also note that the possibility of MAXI~J1957+032 to be a distant source 
contradicts the measured value of extinction via optical observations and $N_\mathrm{H}$.
However, these measurements are based on several assumptions e.g., extinction does not 
significantly vary between active and quiescent states
and if so, then with the current
 data it is not possible to conclusively state whether this system contains a NS or
 a BH.
 
\section{Acknowledgments}
 The authors gratefully acknowledge the referee for his useful suggestions that helped us to improve the 
presentation of the paper.
A.B. gratefully acknowledge the Royal Society and SERB~(Science $\&$
Engineering Research Board, India) for financial support through
Newton-Bhabha Fund. 
A.B. is supported by an INSPIRE Faculty grant (DST/INSPIRE/04/2018/001265) by the Department of Science and Technology, Govt. of India.
She is also grateful to Deepto Chakrabarty (MIT), Peter G.~Jonker (SRON), Craig B Markwardt (NASA/GSFC)
for sharing their \emph{Chandra} data.
DA acknowledges support from the Royal Society.
AP and RW acknowledge support from a NWO Top Grant, Module 1, awarded to RW.
ND is supported by a Vidi grant awarded to ND by the Netherlands Organization for Scientific Research.
We acknowledge the use of public data from the Swift data archive. 
This research has made use of MAXI data provided by RIKEN, JAXA and the MAXI team.

\bibliography{complete-manuscript}{}
 \bibliographystyle{mnras}

   \begin{landscape}
\begin{table}
\def\arraystretch{1.5}%
\caption{Best-fit parameters obtained after fitting the
 spectra (with an absorbed
       	 \textit{bbodyrad+power-law} model)
       	 of two observations that showed a contribution of thermal emission.
Errors quoted are with 90~$\%$ confidence range}
     
 \begin{tabular}{cc |cccccccc}
 \hline
        &              &          &       &    & \\ [0.05ex]

Outburst & Obs~ID & N$_H$ (free) & $kT_{bb}$ & $R_{bb}$  & $\Gamma$ & $N_\mathrm{PL}$ & Total Unabsorbed Flux~(0.5-10) keV & PL Flux/Total Flux & $\chi^{2}_{\nu}$~(dof) \\ [0.5ex] 
&        & 10${^2}{^1}$atoms cm$^{-2}$  &  keV    &  ${km}/4~kpc$,       &    &  $\rm{photons~cm^{-2}~s^{-1}~keV^{-1}}$ & $10^{-11} \rm{ergs~cm^{-2}~s^{-1}}$    &            \\ [0.5ex] 

\hline

Outburst~1 & 00033770001 & $0.02_{-0.02}^{+1.97}$ & $0.42^{+0.03}_{-0.05}$ & $1.9\pm{1.4}$ & $1.3\pm1.0$ & $0.0009_{-0.0003}^{+0.0035}$ & $1.72\pm0.05$ & $0.57\pm0.04$  & 1.28~(34)\\ [0.5ex] 
Outburst~4 & 00033770020 & $0.0_{-0.0}^{+0.8}$ &  $0.41_{-0.04}^{+0.02}$ & $3.4_{-1.7}^{+2.0}$ & $0.42^{+1.17}_{-2.03}$ & $0.0008_{-0.0007}^{+0.0023}$ & $4.8\pm0.2$ & $0.67\pm0.05$  & 0.71~(18) \\ [0.5ex] 

\hline
 \end{tabular}
\vspace{9.4in}
{\bf{Notes}}:~$kT_{bb}$ is the blackbody temperature.~$R_{bb}$ is the blackbody radius in km.~$PL$ stands for power law. \\
            
\label{Best-fit3}  
\end{table}
\end{landscape}

\begin{samepage}
\appendix
\section{Best fit with an absorbed power-law using chi-squared statistics}

Here, we show the spectral fitting using an absorbed power-law model 
of the spectra with good statistics that allowed us to use $\chi^{2}$
statistics to estimate the goodness of the fit.
We grouped the obtained spectra using the ftools task \textsc{grppha} to have at least 25 counts per bin.

 \begin{table}
 \def\arraystretch{1.2}%
\caption{Best fit parameters of MAXI~J1957+032 obtained using the high statistics \emph{Swift}-\textsc{XRT} data.
Errors quoted are for the 90 $\%$ confidence range.~Energy range used is 0.5-10~keV.}
\label{my-label} 
\begin{tabular}{cc |cccc}
 \hline
       &              &          &       &     \\ [0.5ex]       
 
Outburst  & Obs~ID & N$_H$ (free) & $\Gamma$ & $N_{PL}$ & $\chi^{2}_{\nu}$~(dof)  \\ [0.5ex] 

              &        &              &          &            &        \\ [0.5ex]   
             
              &        & (10${^2}{^1}$atoms cm$^{-2}$) &  & ($\rm{photons~cm^{-2}~s^{-1}~keV^{-1}}$) at 1~keV &  \\
\hline

Outburst~1 & 00033770001 & $2.8\pm0.4$ & $2.5\pm0.2$ & $0.007\pm0.001$ & 1.38~(36) \\ [0.5ex] 

\hline
Outburst~2 & 00033770009 & $0.3\pm0.1$ & $2.8\pm0.3$ & $0.010\pm0.002$ & 0.87~(26)  \\ [0.5ex] 
\hline
& 00033770017 & $0.08\pm0.01$ & $1.82\pm0.02$ & $0.176\pm0.005$ & 0.96~(341)  \\ [0.5ex] 
& 00033770018 & $0.07\pm0.01$ & $1.81\pm0.03$ & $0.127\pm0.004$ & 0.89~(262) \\ [0.5ex] 

Outburst~4 & 00033770019 & $0.29\pm0.12$ & $2.1\pm0.3$ & $0.06\pm0.01$ & 0.75~(12) \\ [0.5ex] 

& 00033770020  & $0.27\pm0.07$ & $2.4\pm0.2$ & $0.016\pm0.003$ & 0.85~(20)  \\ [0.5ex]

\hline

\end{tabular}

\end{table}

\begin{figure}
\includegraphics[height=0.75\columnwidth,width=0.75\columnwidth,angle=-90]{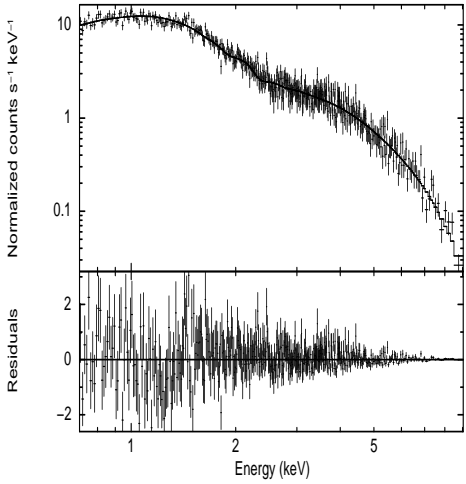}
\caption{This plot shows the residuals obtained using an absorbed power-law model
to the \emph{Swift} observation of MAXI~J1957+032 made on 2016-09-29 during the onset of its brightest outburst in 2017~(outburst~4).}
\label{Outburst-4-chi-square}
\end{figure}
 \end{samepage}   
\end{document}